\documentclass{PoS}

\usepackage[T1]{fontenc}
\usepackage{lmodern}
\usepackage[utf8]{inputenc}
\usepackage{graphicx}
\usepackage{amsmath,amssymb,amsthm}

\newcommand{\be}{\begin{equation}}
\newcommand{\ee}{\end{equation}}
\renewcommand{\d}{\textrm{d}}

\newcommand{\SL}{\mathop{\rm SL}}
\newcommand{\SO}{\mathop{\rm SO}}
 
\title{Instantons, Euclidean wormholes and AdS/CFT}

\ShortTitle{Instantons, Euclidean wormholes and AdS/CFT}

\author{\speaker{Thomas Van Riet}\thanks{Many thanks to my collaborators: T.~Hertog, S.~Katmadas, D.~Ruggeri,  M.~Trigiante and B.~Truijen.}\\
        Instituut voor Theoretische Fysica, KU Leuven,\\
        Celestijnenlaan 200D B-3001 Leuven, Belgium.\\
        E-mail: \email{thomas.vanriet@kuleuven.be}}

\abstract{I present an informal overview of several recent results about Euclidean saddle points sourced by axion fields in quantum gravity (AdS/CFT), such as wormholes, their extremal ``D-instanton" limits and their under-extremal singular counterparts.  Concerning wormholes we argue they cannot contribute to the path integral because a stability analysis suggests they fragment like other super-extremal objects.  For concrete AdS/CFT embeddings the Euclidean saddle point solutions are neatly described by geodesic curves living inside moduli spaces and can typically be solved for using group theory. Our working example is $AdS_5\times S^5/\mathbb{Z}_k$ and allows for smooth Euclidean wormholes. For the supersymmetric D-instanton-like solutions we seem to find a match with the instantons in the dual $\mathcal{N}=2$ quivers.  This match even extends a bit further to self-dual instantons without supersymmetry.  }

\FullConference{Corfu Summer Institute 2019 "School and Workshops on Elementary Particle Physics and Gravity" (CORFU2019)\\
		31 August - 25 September 2019\\
		Corfù, Greece}

\begin{document}

\section{The 3 families of axion-induced Euclidean saddle points}\label{sec1}
When a field theory is Wick rotated to Euclidean signature and contains axion fields, an interesting phenomenon can arise: The Euclidean energy-momentum tensor can source negative Euclidean energy allowing for the famous axion wormhole solutions with finite Euclidean action \cite{Giddings:1987cg, Coleman:1988cy}, see \cite{Hebecker:2018ofv} for a review.

Let us remind ourselves how this works\footnote{This is the only technical part of this proceedings which is there because it is basic and instructive and fits on two pages.}. Consider Euclidean gravity coupled to a free scalar field $\chi$ in $D$ dimensions:
\be\label{action0}
S = -\tfrac{1}{2\kappa^2} \int\sqrt{|g|}\Bigl(\mathcal{R} - \tfrac{1}{2}(\partial\chi)^2\Bigr)\,.
\ee
An action like this is typically a truncation of a larger system. Imagine that this system does not have couplings, at the classical level, that break the shift-symmetry of $\chi$. In that case we can think of the action where the Hodge dual fieldstrength $\star F = \d\chi $ is used instead:
\be
S = -\tfrac{1}{2\kappa^2} \int\sqrt{|g|}\Bigl(\mathcal{R} - \tfrac{1}{2} \tfrac{1}{(D-1)!}F_{\mu_1\ldots \mu_{D-1}}F^{\mu_1\ldots \mu_{D-1}}\Bigr)\,,
\ee
which is classically equivalent to the original action up to a total derivative. A field that can be dualised because of a classical shift symmetry will be named axion field from here onwards. If we plug the expression $F=\star \d \chi$ into the trace-reversed Einstein equation for the field $F$, we find:
\begin{align}
R_{\mu\nu} =& -\tfrac{1}{2}\tfrac{1}{(D-1)!}g_{\mu\nu}F_{\mu_1\ldots \mu_{D-1}}F^{\mu_1\ldots \mu_{D-1}} + \tfrac{1}{2}\tfrac{1}{(D-2)!}F_{\mu \mu_1\ldots \mu_{D-2}}F_{\nu}^{\,\,\,\mu_1\ldots \mu_{D-2}}\,,\nonumber\\
=& \tfrac{1}{2}(-1)^{t+1}\partial_{\mu}\chi\partial_{\nu}\chi\,.
\end{align}
Hence in Euclidean signature $t=0$ we indeed get the (trace-reversed) EM tensor for a scalar with the wrong sign kinetic term. So in the frame where the axion is not dualised one should effectively Wick rotate $\chi\rightarrow i\chi$. Our above explanation is only a shortcut to a slightly more involved explanation first presented in more detail in \cite{Burgess:1989da} (but see also \cite{Bergshoeff:2005zf} or \cite{ArkaniHamed:2007js}). The upshot is that $\chi$ receives an $i$-factor upon Wick rotation depending on the path integral one wishes to compute. If the path integral computes matrix elements where the bra and the ket are axion-charge eigenstates then one indeed has to Wick rotate $\chi$. This we assume from here onwards.

We also consider an extension of the single axion system by allowing multiple fields and a non-zero comological constant, $\Lambda$, which we take either negative (AdS) or zero (Minkowski):
\be\label{action}
S = -\tfrac{1}{2\kappa^2} \int\sqrt{|g|}\Bigl(\mathcal{R} - \tfrac{1}{2}G_{ij}\partial\phi^i\partial\phi^j -\Lambda \Bigr)\,.
\ee
The scalars $\phi^i$ come in two classes: axions and saxions, which is loose language to delineate which fields are wickrotated and which ones not. Hence in Euclidean space the metric $G_{ij}$ on the scalar manifold has mixed signature, and when the space is Lorentzian $G_{ij}$ becomes Euclidean. Note that the possibility to Wick rotate, or to define axions, implies that $G_{ij}$ allows a set of commuting Killing vectors that can be interpreted as shift symmetries. 

We now describe the general spherically symmetric solutions of this action. Our metric Ansatz is given by
\be
\d s^2 = f(\tau)^2\d\tau^2 + a(\tau)^2\d\Omega_{D-1}^2\,, 
\ee
with $\tau$ Euclidean time and $\d\Omega_{D-1}^2$ the line element of the round sphere. If we assume the scalars $\phi^i$ only depend on the radial coordinate the solution for the metric turns out to be universal and independent of the details of the scalar manifold \cite{Breitenlohner:1987dg}. To see this it is useful to observe that the equations of motion for the scalars decouple and are described by the geodesic equation on the scalar manifold with an affine parameter given by the radially symmetric harmonic $h(r)$, defined via $\Box h =0$:
\be
\frac{\d^2}{\d h^2}\phi^i +\Gamma^i_{jk}\frac{\d}{\d h}\phi^j\frac{\d}{\d h}\phi^k=0\,.
\ee
In this parametrisation the scalars have constant velocity
\be
G_{ij}\frac{\d}{\d h}\phi^i\frac{\d}{\d h}\phi^j =  c\,,
\ee
and interestingly the Einstein equations are only sensitive to the constant $c$. Furthermore, the Einstein equations can be simplified to the following constraint equation:
\be
\left(\frac{a'}{f}\right)^2= 1 + \frac{a^2}{l^2} +\frac{c}{2(D-1)(D-2)}a^{-2(D-2)}\,,
\ee
where a prime denotes a derivative wrt to $\tau$ and we introduced the $AdS$ length scale $\ell$ via $\Lambda = -\frac{(D-1)(D-2)}{\ell^2}$.  Note that we can take $\ell\rightarrow \infty$ if we wish to describe asymptotically flat solutions. 
The solution to the above equation can be written in many coordinate frames. An easy frame that allows an explicit expression is obtained by choosing a gauge in which $a=\tau$ and the constraint equation is then trivially solved:
\begin{equation}
\d s^2 = \Bigl( 1+ \frac{\tau^2}{\ell^2} +
\frac{c}{2(D-1)(D-2)}\tau^{-2(D-2)}\Bigr)^{-1} \d \tau^2 + \tau^2\d \Omega^2\,.
\end{equation}
This geometry asymptotes to Euclidean AdS at large $\tau$ regardless of the value of $c$.

\begin{figure}[ht!]
	\begin{center}
		\includegraphics[scale=0.4]{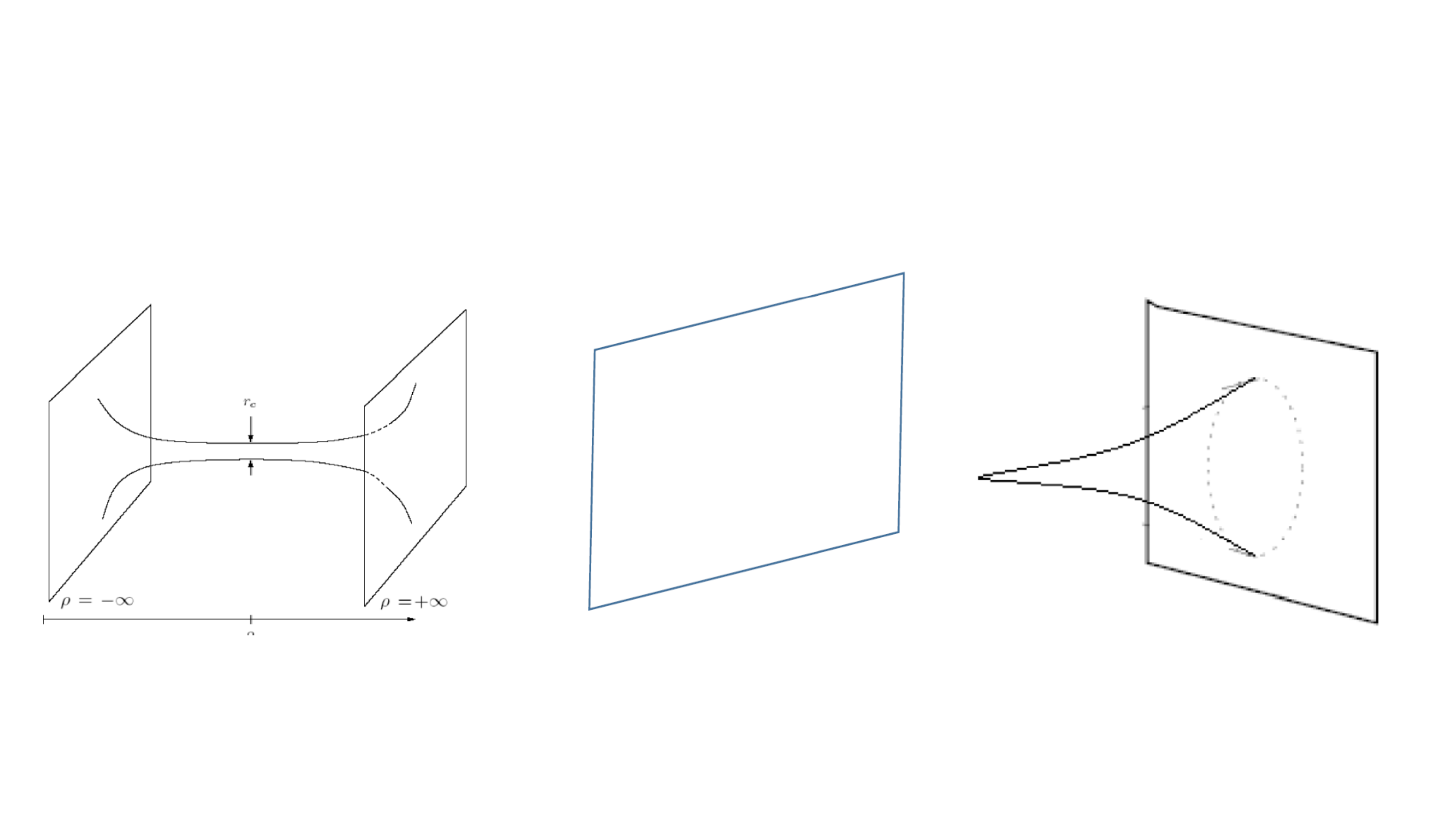}
		\caption{\it The three families of Euclidean geometries from left to right: the wormholes with $c<0$ (timelike geodesics), the extremal D-instanton like solutions with $c=0$ (lightlike geodesics), and the singular deformation of AdS with $c>0$ (spacelike geodesics). }
		\label{Figure1}
	\end{center}
\end{figure}

If there were no axions we would always have $c>0$. But the presence of axions allows the possibility that $c=0$ (lightlike geodesics) and $c<0$ (timelike geodesics). This gives us 3 families of geometries, which are depicted in Figure \ref{Figure1}. 

When $c=0$ we have zero Euclidean energy momentum and the solution to the Einstein equation is simply the vacuum. A well-known example of such solution is the ``D-instanton'' in 10D IIB supergravity which is sourced by the RR axion and the dilaton. This particular solution in 10D is supersymmetric but in general solutions with $c=0$ do not need to be. We rather call them extremal since replacing the single centered harmonic with a multi-centered harmonic still solves the equations of motion (there is a no-force property). When $c<0$ one can show that the metric describes a smooth Euclidean wormhole with a neck radius that scales with $c^{1/2(D-2)}$.\footnote{Although one better uses coordinates in which $f=1$ to see this.} So macroscopic solutions require $c$ to be large enough in Planck units. Finally, when $c>0$ the solutions are singular around the center. 

These 3 geometries (with $\Lambda=0$) first arose in the context of 4d black holes in (ungauged) supergravity \cite{Breitenlohner:1987dg}; if one reduces static black holes over their timelike direction the equations of motion reduce to the system (\ref{action}) with $\Lambda=0$ and can often be solved exactly using group theory. One can show that $c=0$ corresponds to extremal black holes, $c>0$ to under extremal solutions and $c<0$ are the over-extremal solutions with naked singularities. This link with the extremality bound of black holes in 4D has been the source of a renewed interest in axion wormholes as instantons that are required to obey the Weak Gravity Conjecture applied to instantons \cite{Montero:2015ofa, Brown:2015iha, Heidenreich:2015nta, Hebecker:2016dsw}.

We will discuss the physics of each of these 3 classes of solutions in more detail, but we first recall whether the action we have used can be consistently obtained from string theory.

\section{String theory, saxions and AdS moduli spaces}
String theory does not easily give an effective action which would simply equal (\ref{action0}) because axions tend to pair up with saxions. In string vacua of relevance to phenomenology the saxion should be stabilised at high mass scales such that one could potentially integrate out the saxion. However, most backgrounds that are sufficiently well understood do not share this property. So we rather end up with actions of the form (\ref{action}) from truncations of compactified 10D supergravity.\footnote{We refer to \cite{Cortes:2009cs} for an extensive discussion about sigma models in Euclidean supergravity theories with $\Lambda=0$.} The most studied example is a system with a single axion and saxion (\emph{aka} dilaton) forming an $\SL(2,R)/\SO(1,1)$ coset \cite{Gutperle:2002km, Bergshoeff:2004fq, Bergshoeff:2005zf}:
\be\label{axiodil}
G_{ij}\partial\phi^i\partial\phi^j = (\partial\phi)^2 -e^{b\phi}(\partial \chi)^2\,.
\ee
The $\SO(1,1)$ denominator in the coset changes to $\SO(2)$ if there is no Wick rotation. The number $b$ relates to the curvature scale of the scalar coset and depends on the truncation. Let us present the solutions for this particular case in order to be fully explicit for once:
\begin{align}
c<0 \,: \quad&e^{b\phi(\tau)/2} = \frac{|Q|}{\sqrt{|c|}}|\sin\left(\tfrac{1}{2}\sqrt{|c|}b h(\tau)\right)|\,,\quad \chi(\tau) =-\frac{2\sqrt{|c|}}{bQ}\cot\left(\tfrac{1}{2}\sqrt{|c|}b h(\tau)\right) + c_0\,,\nonumber\\ 
c=0 \,: \quad&e^{b\phi(\tau)/2} =\tfrac{1}{2}b|Q| h(\tau)\,,\quad \chi(\tau) = -\frac{4}{b^2Q h(\tau)} +c_0 \,,\nonumber\\
c>0 \,: \quad&e^{b\phi(\tau)/2} = \frac{|Q|}{\sqrt{c}}|\sinh\left(\tfrac{1}{2}\sqrt{c}b h(\tau)\right)| \,,\quad \chi(\tau) = -\frac{2\sqrt{c}}{bQ}\coth\left(\tfrac{1}{2}\sqrt{c}b h(\tau)\right) + c_0\,.\label{solution}
\end{align}
As before $h(\tau)$ is the spherical harmonic on the Euclidean space whereas $Q$ is proportional to the axion charge (coming from the axion shift symmetry which is the origin of the integration constant $c_0$). If we are ignorant about possible singularities we can compute the on-shell actions\footnote{We only present the real part. The imaginary part is each time the same and proportional to the axion charge.} to find:
\be\label{onshell}
S \sim |Q|e^{-b\phi(\infty)/2}\sqrt{1 + \frac{c}{Q^2}e^{b\phi(\infty)}}\,,
\ee
where we dropped overall coefficients coming from the volume of the $S^{D-1}$ and the gravitational coupling.  For the wormhole case we actually only computed the action of the \emph{half} wormhole and assume the contribution from the cut is negligible \cite{Montero:2015ofa, Hebecker:2016dsw}. The action of the full wormhole is the sum of two such terms but with the $\phi$ taking the asymptotic value in the left and the right universe.  In any case the above expression (\ref{onshell}) makes clear why $c<0$ is considered over-extremal, $c=0$ is extremal and $c>0$ under extremal: for instance when $c=0$ we find that tension (action) equals charge times inverse coupling. Surprisingly the wormhole has a lower action than the ``BPS" object ($c=0$). This is a first sign that something strange happens for the wormholes. 

Again, the idea that axion wormholes are ``over-extremal '' makes a close connection with the instanton version of the Weak Gravity Conjecture \cite{Brown:2015iha, Montero:2015ofa, Heidenreich:2015nta, Hebecker:2016dsw}. 

Having these explicit solutions (\ref{solution}) available allows us to discuss singularities in the fields. For $c=0$ and $c>0$ the singularities in the fields are ``standard'' singularities known from p-brane solutions. For $c=0$ there is no metric singularity and whether the metric singularity for $c>0$ is physical will be addressed briefly in section \ref{underextremal}.  For the wormholes ($c<0$) the metric is always regular but regularity of the saxion $\phi$ implies that the range of $h$ from the left to the right of the wormhole should not be too large or otherwise the sin function flips sign. The range of the harmonic depends on the geometry, and for asymptotically flat spacetimes we have a simple formula for regularity \cite{ArkaniHamed:2007js}:
\be\label{regular}
b^2 < \frac{2(D-1)}{D-2}\,.
\ee
For AdS space this formula gets adjusted but with negligibly small terms if the wormhole neck is sufficiently smaller than the AdS length (but still large compaired with the Planck length). In flat space the first example of a compactification giving small enough $b$ was presented in \cite{Bergshoeff:2004pg} where it was shown that small enough $b$ can already be found in a truncation of the universal hypermultiplet generated by compactifying type II string theories on a Calabi-Yau space. For moduli spaces more general than $\SL(2,R)/\SO(1,1)$ one can phrase the regularity criterium in terms of maximal lengths of timelike geodesics \cite{ArkaniHamed:2007js}. Moduli spaces obtained from timelike reductions can never achieve this criterium otherwise there would exist regular over-extremal black holes.

What seems less obvious is whether smooth axion Euclidean wormholes can be embedded into AdS backgrounds of string theory with known holographic duals. A first discussion can be found in \cite{Maldacena:2004rf, ArkaniHamed:2007js} whereas a very concrete embedding was found recently in \cite{Hertog:2017owm}. For this purpose one needs to achieve two things: 1) we need a consistent truncation for which the effective action is given by (\ref{action}) and secondly the metric on the scalar manifold has to be Lorentzian (after Wick rotation to Euclidean AdS) and obey the regularity constraint of \cite{ArkaniHamed:2007js}. Typically consistent truncations of AdS vacua live inside lower-dimensional gauged supergravity theories and when these are truncated to the sector containing scalars and the metric one rather finds effective actions of the form
\be\label{action2}
S = -\tfrac{1}{2\kappa^2} \int\sqrt{|g|}\Bigl(\mathcal{R} - \tfrac{1}{2}G_{\alpha\beta}\partial\phi^{\alpha}\partial\phi^{\beta} -V(\phi) \Bigr)\,,
\ee
where $V(\phi)$ is the scalar potential with AdS minimum. In order to obtain (\ref{action}) we must require that the AdS vacuum has a moduli space that is also a geodesic subspace of the total scalar manifold. If we denote the moduli, ie the scalars that are not fixed by the potential in its minimum as $\phi^i$ where the indices $i$ run over a subset of the indices $\alpha$ we indeed find a consistent truncation of the form (\ref{action}). Interestingly the holographic dual of such AdS moduli $\phi^i$ have the meaning of exact marginal operators in the dual CFT. The set of such operators can often be given a manifold structure and is then named \emph{conformal manifold} and is thought to parametrise a continuous set of CFTs. The metric on moduli space $G_{ij}$ would be dual to the ``Zamolodchikov" metric on the conformal manifold defined through the two-point functions of marginal operators
\be
G_{ij}(\phi)  = y^{2\Delta}\langle \mathcal{O}_i(\vec{y})\mathcal{O}_j(0)\rangle_{S(\phi)} \,,
\ee
where the $\vec{y}$ are ``boundary coordinates" denoting the insertion of the local operator. Marginal operators dual to axion fields are pseudo scalars and can get an $i$ after Wick rotation to Euclidean space and as a consequence the Zamolodchikov metric indeed becomes pseudo-Riemannian, just like the metric of the scalar manifold describing the AdS moduli space. The most well-known example would be an axion that couples to $\text{Tr}F\wedge F$ with $F$ some YM field strength. Then the boundary value of the axion is the theta-angle of the dual YM field theory and indeed the term $\theta \text{Tr}F\wedge F$ is known to get an overal $i$ in the Euclidean formulation.

Let us focus on the well studied case of $AdS_5\times S^5/Z_k$ \cite{Kachru:1998ys}. When $k=1$ the holographic dual is of course $\mathcal{N}=4$ SYM and when $k>1$ the duals are ``necklace" type quivers with $\mathcal{N}=2$ supersymmetry and $k$ gauge nodes \cite{Douglas:1996sw}. The appearance of the conformal manifold in the AdS gauged supergravity has been studied in \cite{Corrado:2002wx, Louis:2015dca}. The moduli space consists of the complex 10D axio-dilaton, as well as $2(k-1)$ real scalars descending from the blow up modes of the orbifold, providing in total $k$ complex scalar fields. These fields are dual to (linear combinations of) the complexified gauge couplings ($1/g_i^2 + i\theta_i $) of the $k$ gauge nodes. These $k$ complex coordinates parametrise a conformal manifold which is a complex hyperbolic space:
\be
\frac{SU(1,k)}{S[U(k) \times U(1)]}\,.
\ee
and the moduli metric takes its canonical form. The $k$ $\theta$-angles manifest themselves by the fact that this coset has exactly $k$ commuting Killing vectors. This makes the Wick rotation of the moduli space unambiguous and one can show it becomes \cite{Hertog:2017owm}
\be\label{lorcoset}
\frac{SU(1,k)}{S[U(k) \times U(1)]}\qquad\rightarrow_{\text{Wick}}\qquad \frac{SL(k+1, R)}{GL(k, R)}\,.
\ee
All the explicit geodesic curves and their properties were studied in \cite{Ruggeri:2017grz, Katmadas:2018ksp}. Such an explicit analysis is possible because geodesics on symmetric coset spaces can be found using the exponential map \cite{Breitenlohner:1987dg}. In the next sections we will discuss the physics of these geodesics according to whether they describe wormholes (timelike geodesics), extremal instantons (lightlike geodesics) or singular under-extremal deformations (spacelike geodesics). But to make the case that wormholes are spurious we stick to the single axion wormhole and just use the above embedding into $AdS_5\times S^5/Z_k$ as a proof of principle that smooth axion wormholes exist within AdS/CFT settings in such a way that all active fields have a very clear holographic interpretation as marginal operators. 

\section{Wormhole fragmentation }
In this section I discuss wormholes assuming the effective action for a single axion field (\ref{action0}) with possibly a negative cosmological constant added. This is for simplicity and could be realised by phenomenological compactifications of string theory with saxions stabilised at high mass. However, as will be clear from our discussion all the intuition should apply straightforwardly to the general case (\ref{action}). The upshot of the discussion is that wormholes do not contribute in the path integral despite the suggestions in the original works \cite{Giddings:1987cg, Coleman:1988cy}.   

 We first start with a subtlety regarding the wormholes as objects with axion charge. Since the wormholes are totally regular they cannot really carry net charge and indeed they carry \emph{dipole} charge. Doing charge measurements around one end of the wormhole would give the opposite charge with respect to the other end. This can be seen in two ways. In the Hodge dual formalism, axion charge is measured by the integral
\be
\int_{S^{D-1}} F_{D-1}\,,
\ee
and the sphere has the opposite orientation on the left with respect to the right universe. Alternatively, one computes the charge using $\partial_h \chi$ and notices that there is a sign flip in the chain rule $dh/d\tau$ once one passes the neck. 

\begin{figure}[ht!]
	\begin{center}
		\includegraphics[scale=0.3]{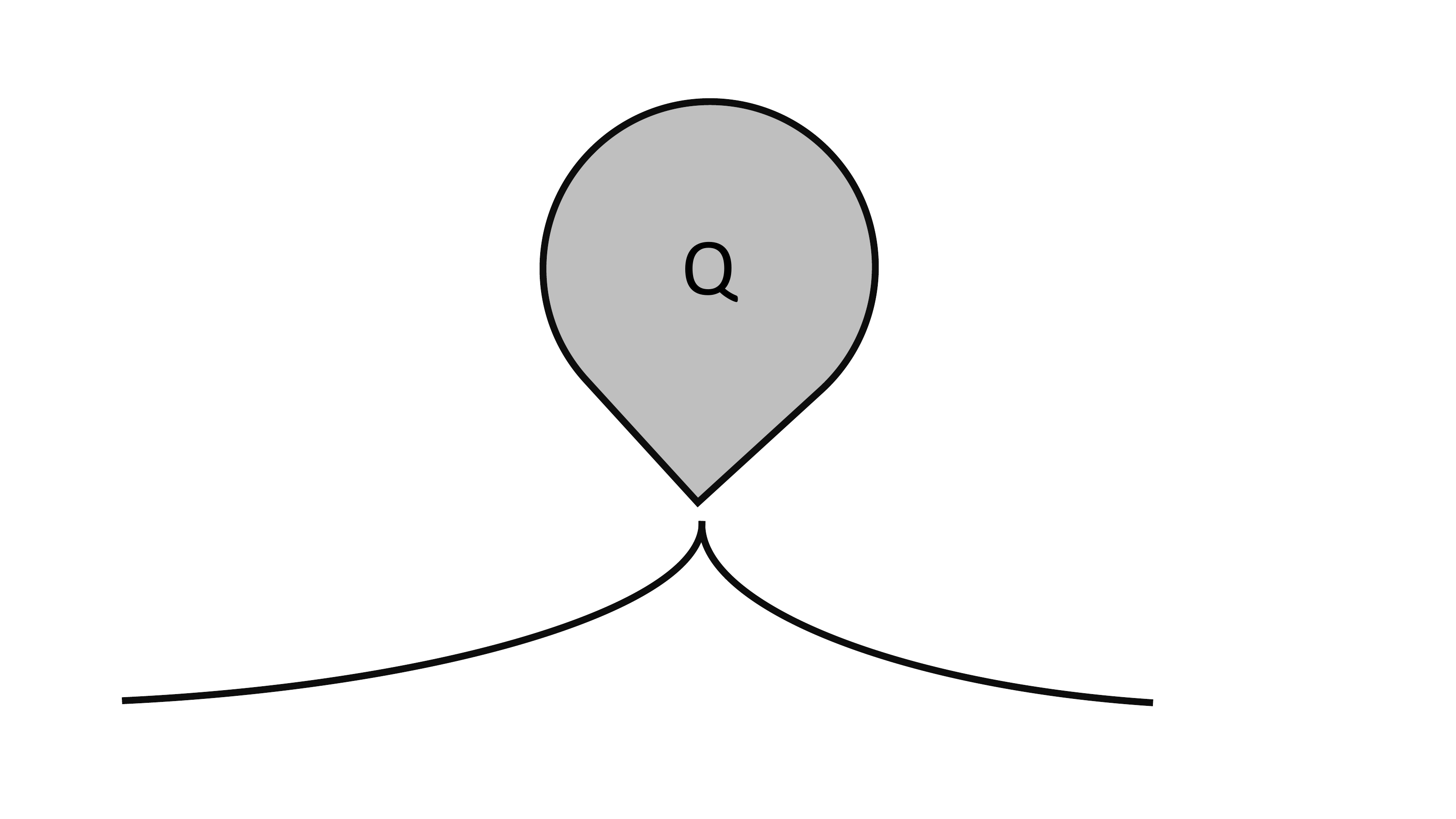}
		\caption{\it An artists impression of the Wick rotation to Lorentzian space of a half wormhole: a baby universe with axion charge $Q$ pinches off (or gets absorbed by) from the mother universe, leading to an apparent violation of axion charge conservation. }
		\label{baby}
	\end{center}
\end{figure}

The most natural interpretation of these wormholes is as instanton-anti instanton pairs where the instanton corresponds to one side of the wormhole that is glued to the other side (anti-instanton) at the neck. The analytic continuation of such a half-wormhole then leads to the interpretation of a tunneling event that creates (or absorbs) a baby universe carrying away (adding) axion charge to the vacuum as depicted in figure \ref{baby}.

At the time this raised some puzzles. First, the violation of axion-charge would lead possibly to measurable quantum incoherence for an observer in the mother universe \cite{Lavrelashvili:1987jg}. Second, the existence of wormhole-like instantons implies a violation of locality of the effective action once a sum over wormhole saddle points is carried out. Coleman, Giddings and Strominger \cite{Coleman:1988cy, Giddings:1987cg} suggested a solution to these problems by interpreting the path integral slightly different and introducing various coupling constants into the effective action drawn from a statistic ensemble. An easy explanation of this using a formal path integral method can be found in the introduction of \cite{ArkaniHamed:2007js}

Holography (from string theory) could provide a concrete test for these ``old school" semi-classical quantum gravity ideas \cite{Maldacena:2004rf, ArkaniHamed:2007js, Betzios:2019rds}. But exactly the property of being over-extremal causes trouble in the holographic context  since the BPS bound in the dual field theories cannot be violated \cite{Bergshoeff:2005zf, Katmadas:2018ksp}. If one ignores the issue of a double boundary or alternatively, one ignores the effect of cutting the wormhole in half, one can perform the usual holographic computation of n-point functions in a wormhole background. This was first attempted in \cite{Bergshoeff:2005zf} for $AdS_5 \times S^5$ and it was found that the vev of the positive operator $(F- \star F)^2$ is negative, signaling a clear contradiction. Strictly speaking the computation in \cite{Bergshoeff:2005zf} was not valid as the wormhole regularity criterion was not satisfied. Nonetheless the explicit regular embedding of \cite{Hertog:2017owm} gives the identical problem in the dual $N=2$ quiver theory \cite{Katmadas:2018ksp}.

Other more general problems also show up in the holographic context. Once the wormholes live inside Euclidean AdS there are two boundaries and hence two CFTs. The CFTs are expected to be decoupled such that the partition functions factorises, but whenever bulk solutions exist that connect the two boundaries the factorisation does not seem to be possible, or obvious, from the gravity side \cite{Maldacena:2004rf, Betzios:2019rds}. It was also pointed out in \cite{ArkaniHamed:2007js} that Coleman's interpretation of varying coupling constants is not supported by AdS/CFT since there is no sign of any $\alpha$-parameters in the dual field theories.

There is some simple intuition which can explain what is happening:  wormholes simply do not contribute to the path integral in the expected way. For this purpose recall that the 3 classes of Euclidean instanton geometries can be found from reducing static black holes in $D+1$ dimensions \cite{Breitenlohner:1987dg}. The wormhole geometries then lift to the unphysical over-extremal solutions. The associated naked singularity in $D+1$ dimensions arises exactly because of the singular saxion profiles discussed around equation (\ref{regular}). This singular saxion is the radion in the timelike reduction and implies that wormholes with smooth fields never correspond to timelike reductions of black holes.  Nonetheless we believe the picture provided by the timelike lift suggests that also smooth wormholes are spurious objects. 

\begin{figure}[ht!]
	\begin{center}
		\includegraphics[scale=0.4]{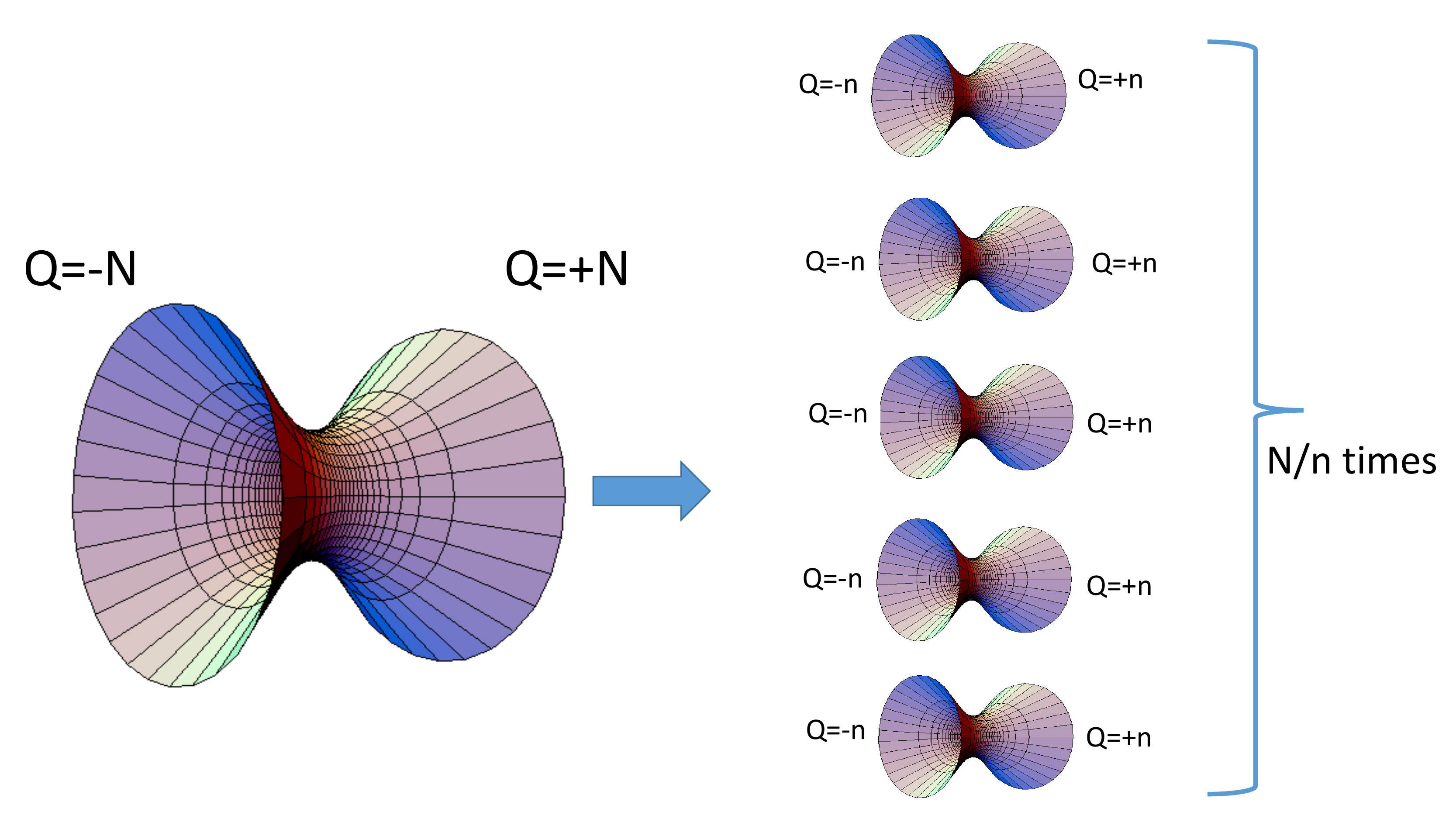}
		\caption{\it On the left we see a wormhole with $N$ units of axion dipole charge going through it. This means $-N$ on the left universe side and $+N$ on the right universe. To the right of this we depict a fragmentation into smaller wormholes with overall the same boundary conditions but charges are now divided over $N/n$ smaller wormholes at a given distance. These separate smaller wormholes also further disintegrate if we search for lower action saddle points, also the distance between all fragments should grow large enough.   }
		\label{wormholefragm}
	\end{center}
\end{figure}

 Clearly macroscopically large objects in Lorentzian space that are over-extremal cannot be formed since any attempt to pile up over-extremal matter would fail and fragment into pieces. It is therefore natural to assume that Euclidean wormholes fragment as well. To understand what this conceptually means one just has to use that the analogue of energy in Lorentzian space is on-shell action in Euclidean space. For instance via timelike reduction black hole mass reduces to instanton action \cite{Breitenlohner:1987dg}. Over-extremal objects brought together have a higher energy than if they would be at far distance. That is why it fragments. Similarly, we expect the on-shell action of fragmented wormholes to be smaller than the on-shell action of the ``larger" wormhole as depicted in Figure \ref{wormholefragm}.
If two wormholes ``repel" in a Euclidean sense it means that the action of two wormholes gets smaller by separating them. To compute the binding energy between wormholes is not immediately straightforward, but what we can check is what happens to the action of ``probe" D-instantons near wormholes. If probe D-instantons get repelled then it makes a good case that wormholes get repelled since, at least from the viewpoint of a specific side of the wormhole it corresponds to an object with a higher charge/tension ratio. The probe action for a (anti-)D-instanton is
\be
S_{probe} = e^{-b\phi/2} \pm \frac{b}{2}\chi   \,.
\ee
When inserted in the extremal background we find the action is constant as expected (if we take the right charge). This explains why multi-centered harmonics also solve the equations of motion with the same Ansatz. When we instead put probe D-instantons in the non-extremal backgrounds we find the following actions for them
\begin{align}
c>0\, : \quad & S_{probe} = \frac{\sqrt{c}}{|Q|}\frac{1}{\sinh(\tfrac{1}{2}\sqrt{c}b h(\tau))}\left(1-\cosh(\tfrac{1}{2}\sqrt{c}b h(\tau))\right)\,,\\
c<0\, : \quad & S_{probe} = \frac{\sqrt{|c|}}{|Q|}\frac{1}{\sin(\tfrac{1}{2}\sqrt{|c|}b h(\tau))}\left(1-\cos(\tfrac{1}{2}\sqrt{|c|}b h(\tau))\right)\,.
\end{align}
In our choice of coordinates spatial infinity correspond to $h=0$ and larger values of $h$ means moving towards the center of the space. Then we can easily deduce that for wormholes ($c<0$) the action is minimized at spatial infinity and the opposite for the sub-extremal geometry ($c>0$). So if wormholes repel D-instantons they will certainly repel each other. This does not show that wormholes are perturbatively unstable but it is nonetheless a sign that they can be. Since perturbative instability in the Euclidean sense tends to imply there are other configurations with the same boundary conditions that have a lower action. We claim these are the fragmented wormholes, which in turn fragment separately etc, to get a lower action.

A computation suggesting the wormholes are unstable perturbatively was carried out in \cite{Hertog:2018kbz} by computing the quadratic action, revising existing results \cite{Rubakov:1996cn, Kim:2003js, Alonso:2017avz}.  Recall that saddle point expansions work as follows:
\be
Z \approx e^{-S[\Phi_0]}\int D\varphi e^{-\delta^2 S[\Phi_0,\varphi]+\mathcal{O}(\varphi^3)} \,.
\ee
Here $Z$ denotes the partition function with certain boundary conditions, $S$ the classical action, $\Phi_0$ collectively denotes the on-shell background field configurations and $\varphi$ is shorthand for the field fluctuations around the classical background. One can then write the quadratic part of the action symbolically as
\be
\delta^2 S = \frac{1}{2}\int \varphi \hat{M} \varphi\,,
\ee
with $\hat{M}$ the quadratic operator. If $\hat{M}$ can be diagonalised 
\be
\frac{1}{X}\hat{M}\varphi_n =\lambda_n \varphi_n\,,\qquad \int X \varphi_n\varphi_m=\delta_{mn} \,,
\ee
with $X$ some weighting function chosen for convenience,
we find
\be
Z = e^{-S[\Phi_0]}\frac{1}{\sqrt{\det(\hat{M})}}\,, 
\ee
where the determinant is some regularisation of $\Pi_n\lambda_n$\,. As pointed out by Coleman long time ago \cite{Coleman:1987rm}, if there is no negative eigenvalue the instanton contributes to quantities like energies in the form of a potential etc. If there is just one negative mode the instanton rather contributes imaginary values to the energy which can be interpreted as the usual tunneling effect in quantum mechanics. If there are multiple negative modes the instanton does not contribute. See \cite{Bramberger:2019mkv} for a recent discussion on this in the context of theories with gravity.

In reference \cite{Hertog:2018kbz} this stability computation was carried out using gauge invariant variables and with correct boundary.  It was found that the quadratic operator was not diagonalisable but it nonetheless shows the wormholes cannot contribute. The reason is that the quadratic action allowed \emph{infinitely} many normalisable modes which obeyed proper boundary conditions and nonetheless lower the action:
\be
\int \varphi \hat{M}\varphi <0\,.
\ee
Crucually all these modes were concentrated around the neck region of the wormhole and all modes that do not probe the wormhole topology give positive contributions. The reader worried about the fact that the quadratic operator in \cite{Hertog:2018kbz} is not obviously diagonalisable should take comfort in the following observations strengthening the case made in \cite{Hertog:2018kbz}:
\begin{itemize}
	\item If an instanton is spurious because of a multitude of fluctuations that can lower the action, one does not require to compute any determinant in the first place. So there is no need to diagonalise the operator. 
	\item The quadratic action found in \cite{Hertog:2018kbz} does \emph{not suffer from the conformal factor problem} that otherwise haunts many works on semi-classical gravity, including the previous works \cite{Rubakov:1996cn, Alonso:2017avz} on wormhole stability that had to artificially introduce the Hawking-Perry analytic continuation.  The reason there is no conformal factor problem is due to the absence of a homogenous mode once correct boundary conditions are imposed. In fact this had to since the analysis in section \ref{sec1} shows clearly that once the axion charge is fixed the metric is solved from a constraint equation alone. All the remaining freedom is gauge transformations. 
	\item Once the correct boundary conditions were enforced and the gauge invariant variable chosen one arrives at a quadratic action with no singularities in the kinetic term. This is a very healthy feature since a typical quadratic action one obtains from some arbitrary set of boundary conditions (that are all related by symplectic rotations) almost always contains divergent kinetic terms and this obscures the eigenmode problem. Typically one has to fine-tune the ``sympletic frame" in order to find a regular operator \cite{Alonso:2017avz}. We found a regular operator by insisting to fixed axion charge in the path integral  \cite{Coleman:1988cy, Giddings:1987cg, Burgess:1989da}. In practice this means a symplectic frame in which axion momentum is fixed at the boundary.
\end{itemize}

Finally let us mention that the mode analysis of \cite{Hertog:2018kbz} is consistent with the intuitive picture of wormhole fragmentation, or wormhole repulsion. If the wormhole fragments into smaller pieces in order to lower Euclidean energy, it should do so by activating the non-homogenous modes because the fragmentation obviously destroys the spherical symmetry.

If wormhole fragmentation indeed is the correct picture, the leading saddle points should have microscopic neck size and hence be Planckian. This means that the instanton expansion is not under control and there is no notion of macroscopically sized wormholes contributing in the path integral. 

\section{Extremal instantons and self-dual YM instantons}
In what follows we move to the geometries with $c=0$ as embedded in $AdS_5\times S^5 /\mathbb{Z}_k$ \cite{Ruggeri:2017grz}. So the spacetime geometry is given by undeformed Euclidean $AdS_5\times S^5 \mathbb{Z}_k$ since the energy momentum of the scalars is such that the negative axion contributions exactly cancel the positive saxion contributions. The scalars are described by lightlike geodesics on the manifold (\ref{lorcoset}) and they can be solved exactly using the exponential map. The corresponding Noether charge matrix $\mathcal{Q}$ appearing in the exponential sits in the coset algebra and is nilpotent of degree two or three (in the coset representation used in \cite{Ruggeri:2017grz})
\be
\text{lightlike geodesics:}\qquad \mathcal{Q}^2=0\quad \text{or}\quad \mathcal{Q}^3=0\,.
\ee
Elegantly, supersymmetry turns out equivalent to $\mathcal{Q}^2=0$ \cite{Ruggeri:2017grz} and  solutions with $\mathcal{Q}^3=0$ are extremal but non-SUSY. Note that $\mathcal{Q}^3=0$ requires $k>1$ and so does not exist in $AdS_5\times S^5$ (denoted $k=1$ from here onwards). In fact for $k=1$ the solution is well known and corresponds to the insertion of a $D(-1)$ brane inside $AdS_5\times S^5$ or alternatively, it is the near horizon limit of the $D(-1)/D3$ bound state. This case is very well studied in the early days of AdS/CFT and it was verified that these supergravity instantons indeed map to the SUSY instantons of $N=4$ SYM as reviewed in \cite{Belitsky:2000ws}. The easiest match one can do is to map the following items in the holographic dictionary:
\begin{enumerate}
\item Zero-modes (fermionic and bosonic). 
\item On-shell actions (real and imaginary parts).
\item The vevs of Tr$F^2$ and Tr$F\wedge F$.
\end{enumerate}
And obviously the amount of preserved supergenerators ($Q$'s and $S$'s.) 
Note that the AdS moduli space has exactly $k$ commuting shift symmetries whose associated quantised charges correspond to the $k$ different Potryagin indices (winding numbers) of the dual gauge nodes. Items 2 and 3 above have been explicitly checked to work out for the SUSY instantons in $AdS_5\times S^5 /\mathbb{Z}_k$. It would be interesting to perform further checks of the correspondence. However, let us remark that the whole class of SUSY instantons lies on a ``duality-orbit" of the $D(-1)$ solution inserted into $AdS_5\times S^5/ \mathbb{Z}_k$ \cite{Ruggeri:2017grz} and we therefore do expect this match to arise. Although this is only intuition at this point, not more. 

\begin{figure}[ht!]
	\begin{center}
		\includegraphics[scale=0.3]{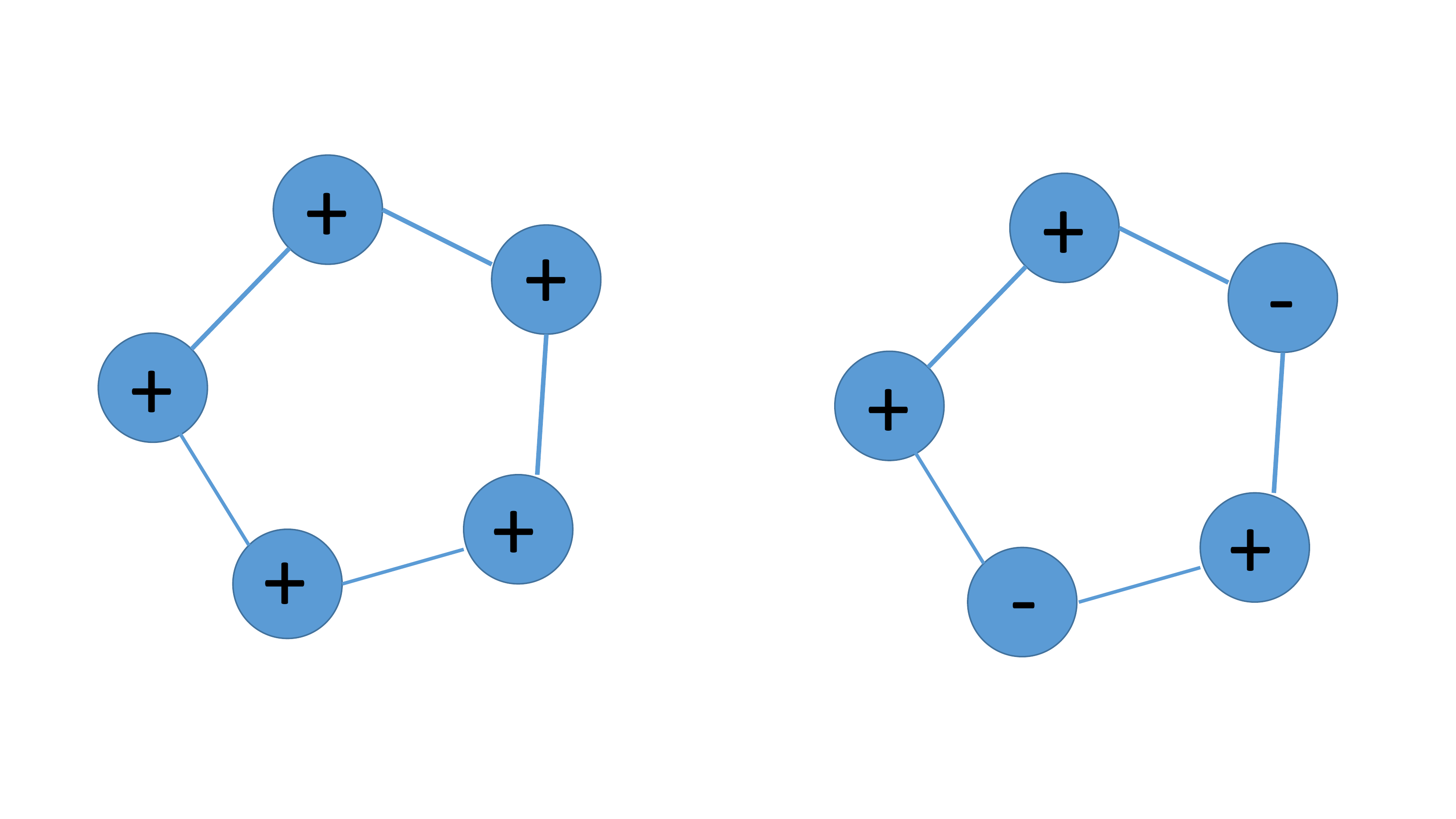}
		\caption{\it On the left we depict a quiver with 5 nodes where at each node we have a self-dual instanton with certain winding number. On the right we have winding numbers of different signs (ie self-dual and anti-self-dual field strengths). Such configurations correspond to the quasi-instantons mentioned in \cite{Imaanpur:2008jd}.}
		\label{quiver}
	\end{center}
\end{figure}

Concerning the solutions with  $\mathcal{Q}^3=0$ we found that a subclass of them seem to holographically match the so-called ``quasi instantons" mentioned in \cite{Imaanpur:2008jd}.  Quasi instantons are objects that arise in YM field theories with several gauge nodes, exactly as in our $N=2$ quiver theories dual to IIB string theory in  $AdS_5\times S^5 /\mathbb{Z}_k$. SUSY instantons are such that the instanton number on each gauge node has the same sign, such that the gauge field winding is in the same orientation. Quasi instantons solve the Yang Mills equations and are (anti-) self-dual on each separate gauge node. But they fail to be supersymmetric since each gauge node preserves a different part of the 4D SUSY generators. We symbolically represent the difference between SUSY and quasi instantons in figure \ref{quiver}.

The holographic ``match" in \cite{Katmadas:2018ksp} only involved the computation of the one-point functions, the quantisation of charges, and the on-shell actions, athough there is a non-trivial subtlety if one wishes to match the on-shell actions \cite{Katmadas:2018ksp}.

We cannot help to observe that AdS/CFT seems to provide us with a  map, which becomes rather remarkable once rephrased in pure field theory language:\\

\emph{There is a correspondence between lightlike geodesics on the conformal manifold of a CFT at large $N$ and the instanton configurations of that same CFT.}\\

It would be interesting to understand whether there is something deeper to this statement, because a priori there is no obvious relation between conformal manifolds and instantons.

\section{Singular ``under-extremal" solutions}\label{underextremal}
The solutions with $c>0$ are the usual equivalent of non-extremal extensions of p-branes but with $p=-1$.  The geometry has a spikelike singularity in the middle of space (Euclidean flat space or AdS) and it is difficult to interpret it. For solutions coming from timelike reductions we can, since the solution lifts to non-extremal black holes which are physical. However, for genuine instantons we are in the dark. One worrisome thought is that these solutions can be regarded as the Euclidean analogues of Gubsers "dilaton driven confinement"-flow \cite{Gubser:1999pk} whose singular behavior might be unphysical \cite{Gubser:2000nd}. To understand this link notice that spacelike geodesics ($c>0$) on $\SL(2,R)/\SO(1,1)$ can always be rotated with an $\SL(2,R)$ symmetry to a solution in which the axion is constant. This can be then be trivially Wick rotated to Lorentzian AdS and one obtains the solution in \cite{Gubser:1999pk}.

However, the above is still unclear and in fact a suggestion of a possible holographic dual was made in \cite{Bergshoeff:2005zf} for $k=1$ and extends trivially to $k>1$ \cite{Katmadas:2018ksp}. It is in spirit of the quasi instantons of \cite{Imaanpur:2008jd}. Consider, in the large $N$ limit the color space matrix corresponding to $F_{\mu\nu}$
\be
F_{\mu\nu}^{\rm SU(N)}=\begin{pmatrix}
	F_{\mu\nu}^{\rm SU(2)} & 0 & \ldots & 0\\
	0 & F_{\mu\nu}^{\rm SU(2)} &  & 0\\
	\vdots  &  & \ddots  & \\
	0   & &  & \overline{F}_{\mu\nu}^{\rm SU(2)}
\end{pmatrix}\,.
\ee     
 On its diagonal there are various mutually commuting $SU(2)$ factors. One trivial set of SUSY instantons is when these separate $SU(2)$ factors all have the same orientation as (anti) self-dual $SU(2)$ solutions. In fact the ADHM construction is concerned with counting how many instantons are related to these basic building blocks. One trivial method to still solve the Euclidean classical YM equations, but breaking (anti-) self-duality of the $SU(N)$ field strength is by ``doping it'' with an opposite orientation $SU(2)$ block, denoted with $\bar{F}$ in the above equation. 

So unlike the quasi instantons of the previous sections these solutions are neither self-dual or anti-self-dual in a separate gauge node. Of course whether these configurations really are described by the non-extremal supergravity instantons is a long shot, but if they have a weakly coupled dual, it would be strange they are not the non-extremal supergravity solutions. 

\section{Outlook}
I have overviewed in a rather conceptual manner the main results behind the papers \cite{Hertog:2017owm, Hertog:2018kbz,Ruggeri:2017grz,Katmadas:2018ksp}. 

One of the most important results we overviewed and expanded upon, is the suggestion from \cite{Hertog:2018kbz} that Euclidean axion wormholes do not contribute to the path integral. If so axion wormholes would be spurious objects much in line with the property that they can be regarded as over-extremal objects.  Whereas the unphysical nature of over-extremal ``black" holes are clearly flagged by their blatant violation of cosmic censorship, smooth Euclidean wormholes seem to be rather innocent in that respect.\footnote{Although we already overviewed the laundry list of conceptual issues they bring with them.}
We find it rather surprising that the possibility that the wormholes do not contribute is seldom considered, but nonetheless is not in contradiction with a recent critique on our understanding of baby universes \cite{1791270} or a recently noticed tension between the calculus of replica wormholes versus axion wormholes \cite{Giddings:2020yes}. Or as stressed in \cite{ArkaniHamed:2007js}; embeddings of axion wormholes in AdS/CFT pairs do not show any sign of Coleman's $\alpha$-parameters or stochastic coupling constants. 

Motivated by the search for smooth Euclidean wormholes in established AdS/CFT dual pairs we put much emphasis on the $AdS_5\times S^5/\mathbb{Z}_k$ background of IIB string theory. The Euclidean supergravity solutions (instantons) with axion charge then correspond neatly to geodesic curves on the manifold (\ref{lorcoset}) and can be explicitly solved. The wormholes are timelike geodesics and can be smooth. We furthermore investigated the lightlike and spacelike geodesics. Especially in the case of lightlike geodesics we found a nice match between properties of (anti-)self-dual field theory instantons with Potryagin indices $N_i$ (where $i$ runs over the $k$ gauge nodes) and the quantised axion charges $N_i$. Supersymmetry required all $N_i$'s to have the same sign, but it seems supergravity can also describe (to some extent) the non-SUSY ``quasi-instantons" that have different signs for the various $N_i$. 

It appears that Euclidean supersymmetric flows inside AdS moduli spaces are rather unexplored territory when compared with domain wall flows that connect different AdS vacua that are dual to RG flows. At least from the supergravity viewpoint of classifying supersymmetric trajectories of gauged supergravities, the instanton flows inside the moduli space should be equally relevant and mathematically rich.

Many of the things I reported on can be understood much better and below I provide a list with problems that could be of interest to tackle:
\begin{itemize}
	\item The wormhole stability analysis of \cite{Hertog:2018kbz} crucially relies on a gauge invariant formalism in which the fluctuations of the axion are combined with scalar perturbations of the metric into one gauge invariant variable  $\mathcal{X}$. The quadratic action of the modes in $\mathcal{X}$ is then found by Wick rotating the Lorentzian action and giving $\mathcal{X}$ and $i$-factor. But since $\mathcal{X}$ is not purely the axion scalar one could worry about this procedure. An unambiguous method relies on perturbation theory in the Hodge dual frame with form fields instead. An investigation of this is under way \cite{Simon}.
	\item The wormhole stability analysis should be extended to the multi-field system (\ref{action}) with saxions and axions. We believe this is doable in full generality since the background solutions for the scalars are geodesic and as a consequence geodesic coordinates simplify greatly the computations.
	\item It would be interesting to verify whether the stability of the half-wormholes is the same as the full wormholes.
	
	\item If wormholes fragment it should be possible to check this without a full blown stability analysis but by computing the ``Euclidean force" between two wormholes to verify whether they repel.
	
	\item It would also be interesting to verify how Euclidean axion wormholes are corrected at distances smaller than the inverse mass-scale of the stabilised saxion. This would provide a classical, yet stringy correction to ``small" wormhole solutions \cite{Brecht}.
	
	\item The link between lightlike geodesics on conformal manifolds and instanton solutions of a CFT is quite intriguing \cite{Ruggeri:2017grz, Katmadas:2018ksp}. It would be interesting to find other examples of this correspondence. At the supergravity level it would for instance be doable to find the supersymmetric and non-supersymmetric geodesics on the moduli-space of $AdS_3\times S^3\times T^4$ (or $\times K^3$). From the CFT perspective \cite{Cecotti:1990kz} one expects this moduli space to be (a Wick rotation of) the coset: $SO(n,4)/SO(n)\times SO(4)$ with $n=4$ for $T^4$ and $n=20$ for $K^3$, although this has not yet been verified in supergravity. 
\end{itemize}

\section*{Acknowledgements}

It is a pleasure to thank the organisers of "School and Workshops on Elementary Particle Physics and Gravity" (CORFU2019) and my dear collaborators Thomas Hertog, Stefanos Katmadas, Daniele Ruggeri, Mario Trigiante and Brecht Truijen. Not all viewpoints of this paper might be shared by them. I would also like to thank Manuel Kr\"amer and my master students Simon Maenout and Brecht Wagemans.  Finally I would like to thank Miguel ``Migwel" Montero and Pablo ``telebancos" Soler for many discussions about wormholes which have shaped my understanding of the subject. My work is supported by the KU Leuven C1 grant ZKD1118 C16/16/005.

\end{document}